\documentclass[10pt,prd,aps,superscriptaddress,showpacs]{revtex4-1}
\usepackage{graphicx}
\usepackage{epsfig}
\usepackage{amssymb}
\usepackage{amsmath}

\def\plotone#1{\centering\leavevmode\epsfxsize=.95\textwidth\epsfbox{#1}}

\begin{document}
\title{The Matter Power Spectrum of Dark Energy Models
and the Harrison-Zel'dovich Prescription}
\author{Ivan Duran \footnote{E-mail address: ivan.duran@uab.es}}
\affiliation{Departamento de F\'{\i}sica, Universidad Aut\'{o}noma de Barcelona,
Barcelona, Spain}
\author{Fernando Atrio-Barandela\footnote{E-mail address: atrio@usal.es}}
\affiliation{Departamento de F\'{\i}sica Fundamental, Universidad
de Salamanca, Spain}
\author{Diego Pav\'{o}n\footnote{E-mail address: diego.pavon@uab.es}}
\affiliation{Departamento de F\'{\i}sica, Universidad Aut\'{o}noma de Barcelona,
Barcelona, Spain}

\begin{abstract}
According to the Harrison-Zel'dovich prescription, the amplitude
of matter density perturbations at horizon crossing is the same at
all scales. Based on this prescription, we show how to construct
the matter power spectrum of generic dark energy models from
the power spectrum of a $\Lambda$CDM model without the need of
solving in full the dynamical equations describing the evolution
of all energy density perturbations. Our approach allows to make
model predictions of observables that can be expressed in terms of
the matter power spectrum alone, such as the amplitude of matter
fluctuations, peculiar velocities, cosmic microwave background
temperature anisotropies on large angular scales
or the weak lensing convergence spectrum.
Then, models that have been tested only at the background level
using the rate of the expansion of the Universe can now be tested
using data on gravitational clustering and on large scale
structure. This method can save a lot of effort in checking the
validity of dark energy models. As an example of the accurateness of the
approximation used, we compute the power spectrum of different dark energy
models with constant equation of state parameter ($w_{DE}=-0.1$, $-0.5$ and $-0.8$,
ruled out by observations but easy to compare to numerical solutions)
using our methodology and discuss the constraints imposed by the low
multipoles of the cosmic microwave background.

\end{abstract}

\maketitle

\section{Introduction}\label{sec:Introduction}
In the concordance $\Lambda$CDM cosmological model, the current
accelerated phase of expansion is driven by a cosmological
constant $\Lambda$ that dominates the present energy density of
the Universe, whose equation of state (EoS) parameter is $w=-1$.
The second component in importance is cold dark matter (DM), a
non-baryonic dust component that drives the growth of large scale
structure (LSS). This simple model fits rather well the
observational data and requires the minimum set of cosmological
parameters \cite{komatsu}. Also, it is the preferred model based
on statistical selection criteria \cite{liddle_04}. The
observational successes of the $\Lambda$CDM cosmology is linked to
its capacity to reproduce the right sequence of cosmological eras:
matter-radiation equality occurs well before recombination and the
matter dominated period lasts long enough to allow the growth of
LSS. The length of the radiation and matter periods are crucial to
determine the shape of the matter and radiation power spectra.
Alternative models must also reproduce the correct sequence of
cosmological eras to fit the data \cite{amendola_06}.

While the $\Lambda$CDM model fits the observational data very well, it
appears rather unsatisfactory from the theoretical point of view.
In fact, a cosmological constant is not very appealing. Its
measured value is 120 orders of magnitude smaller than the
expected amplitude at the Planck scale \cite{weinberg_89} and
introduces the so-called coincidence problem (i.e., ``why are the
densities of DM and dark energy (DE) of the same order precisely today?")
\cite{steinhardt}. This is why a plethora of more flexible models
that behave akin to the $\Lambda$CDM at the background level have
been introduced over the years -see \cite{recent-reviews} for
recent reviews. This complicates enormously the task of
judiciously deciding which model should be preferred over all the
others in view of their observational and theoretical merits. For
instance, the acceleration could be driven by a dark energy
component with EoS parameter $w_{DE}\neq -1$ \cite{peebles-ratra},
constant or variable on cosmological timescales. The simplest
variants require cosmological parameters to be fine tuned at some
initial time, suffering also -though, at a lower extent- from the
coincidence problem  and more complex models have been introduced
\cite{olivares}. Before carrying out a detailed analysis, these
alternatives first use probes of the cosmic expansion history such
as luminosity distances derived from supernovae type Ia data,
angular diameter distances from baryon acoustic oscillations (BAO), the
expansion rate, $H$, at various redshifts, etc.
\cite{lazkoz,ivan-pavon}. Data on matter density perturbations and
cosmic microwave background (CMB) anisotropies provide stronger
constraints but require to solve the time evolution of the density
perturbations in all components. In general, the resulting set of
equations is far more involved than in the standard $\Lambda$CDM
model. Furthermore, small differences on the dynamics of the dark
sector change the equations governing the evolution of matter
perturbations \cite{wang}, and no generic constraints can be
imposed on large classes of models. As a result, many models in
the literature have not been constrained by the current data on
density inhomogeneities and CMB temperature anisotropies -see e.g. \cite{blevel}.

The aim of this paper is to show how to derive the matter power
spectrum of a generic DE model without the need of solving in full
the perturbation equations for the radiation and matter
components. Several observables can be computed in terms of the
matter power spectrum alone and can be used to constrain the
model beyond the expansion rate. The observables include:
(a) the fluctuation of the matter density perturbations on a
sphere of $8h^{-1}$Mpc, $\sigma_8$ \cite{kash-review}; (b) the rms
peculiar velocity of matter on spheres of radius $R$, $\langle
v^2(R) \rangle^{1/2}$ \cite{strauss-willick}; (c) the weak lensing
convergence spectrum \cite{weak-lensing}; (d) the Sachs-Wolfe (SW)
and Integrated Sachs-Wolfe (ISW) components of the CMB temperature
anisotropies \cite{sachs-wolfe}; (e) the cross-correlation of the
ISW with templates of projected density of galaxies \cite{isw},
etc. For example, the SW and ISW  effects are the dominant
contributions to the CMB anisotropies at low multipoles. If the
power spectrum is normalized to the measured value
$\sigma_8=0.801\pm 0.030$ \cite{larson}, the predicted low order
multipoles of the CMB, the peculiar velocity on a given scale
\cite{kabke} or the measured ISW-Large Scale Structure
cross-correlation \cite{isw-olivares} can be compared with
observations. Our method provides simple tests of models using a
wealth of data beyond luminosity and angular diameter distance
measurements.

The idea of how to construct the matter power spectrum is based on
the Harrison-Zel'dovich (HZ) \cite{harrison-zeldovich}
prescription. For any two models, the amplitude of density
perturbations are specified at horizon crossing instead of at some
arbitrary initial hypersurface. The final spectrum will differ
only by the subsequent (subhorizon) evolution of each single
mode. If we use as a starting model one that is implemented on
publicly available numerical codes like CMBFAST \cite{cmbfast},
for example the concordance $\Lambda$CDM model,
we will be able to construct the power spectrum of a generic
DE model (not implemented on numerical codes).
The method can be used to derive the power spectrum ranging
from galaxy to horizon scales, i.e., at all the scales that can be
observed at present.

In developing this method, our aim is not to use it to solve models of
DE with a constant EoS parameter. In section \ref{sec:MPS_DE}, we consider
a DE model with constant EoS parameter, $w=-0.5$ (though being aware that it
is observationally discarded) just to illustrate the accurateness of the method
in a model which, on the one
hand, it is easy to obtain the exact evolution of the perturbations and,
on the other hand, its evolution  differs substantially from that of the
standard $\Lambda$CDM. With this we can see that obtaining the matter
power spectrum of a DE model from the one of $\Lambda$CDM
is reasonable also in cases when both models differ greatly at the background level.

This method can be useful in solving the perturbation equations of
interacting DE-DM models \cite{Chimento, Wang:2005jx,
Pavon:2005yx, ivan-pavon}. Many of these models aim to describe
the Universe at low redshifts, when DM and DE dominate the
expansion. Following our method, the matter perturbations evolve
as in the $\Lambda$CDM model before horizon crossing (this avoids
the need of explicitly introducing initial conditions for the
perturbations). Moreover, because it suffices to compute the
evolution of the perturbations of the particular DE model
considered after they enter the horizon, the set of equations to
be solved gets greatly simplified. This is shown in section
\ref{sec:MPS_DE}.

Briefly, section \ref{sec:HZP} recalls the basics of the HZ
prescription. Section \ref{sec:MPS_DE} describes how the matter
power spectrum of a generic DE model can be constructed. Section
\ref{sec:RPS} discusses how to calculate the low multipoles on CMB
temperature anisotropies to constrain the model. Finally, Section
\ref{sec:Conclusions} summarizes the main results of the paper.

\section{The Harrison-Zeldovich prescription}\label{sec:HZP}
In their seminal papers, Harrison and Zel'dovich
\cite{harrison-zeldovich} computed the present matter power
spectrum assuming that all perturbations at horizon crossing have
the same amplitude. Then, they computed the matter power spectrum
$P(k)$ at the present time after accounting for the subhorizon
evolution of each mode. To illustrate their argument, let us
construct the power spectrum of the standard CDM model, a model
that only contains DM, baryons and radiation (subscripts $c$, $b$,
and $r$, respectively), and verifies
$\Omega_{c}+\Omega_b+\Omega_{r}=1$, where $\Omega_i$ is the energy
density of component $i$ in units of the critical density. Let us
define the density contrast by
$\delta(k,t)=(\delta\rho/\bar{\rho})(k,t)$. In the linear regime,
spatial and time dependence can be separated:
$\delta(k,t)=\delta(k)D_+(t)/D_+(t_0)$, where $D_+(t)$ denotes the
growing solution and $\delta(k)$ is evaluated at the present time,
$t_0$. The current power spectrum is then defined as:
$P(k)=|\delta(k)|^2$.

In the standard cold dark matter (CDM) model, $D_+(t)\approx$
const during the radiation dominated era and
$D_+(t)=D_+(t_{in})(t/t_{in})^{2/3}$, during the matter dominated
period. The HZ prescription establishes that all mass
perturbations, defined as $\Delta(k,t)=\frac{1}{2\pi^{2}}k^3
P(k)(D_+(t)/D_+(t_0))^2$, have the same amplitude at the time
$t_{in}$ when they enter the horizon, i.e.,
$a(t_{in})\lambda_{in}=d_H(t_{in})$ where $a(t)$ is the scale
factor, $d_H(t)$ the radius of the horizon, and $\lambda_{in}$ the
comoving wavelength of each particular mode;
$k_{in}=2\pi/\lambda_{in}$ would be the corresponding wavenumber.
In particular
\begin{equation}
\Delta(k_{in},t_{in})= {\rm const} =\Delta(k_{eq},t_{eq}) ,
\end{equation}
with $t_{eq}$ the moment of matter-radiation equality.

Once a perturbation enters the horizon, it will evolve as $D_+(t)$ and we can write
\begin{equation}
\Delta(k_{eq},t_{0})=\left(\frac{D_+(t_0)}{D_+(t_{eq})}\right)^2\Delta(k_{eq},t_{eq})
=\left(\frac{D_+(t_0)}{D_+(t_{eq})}\right)^2\Delta(k_{in},t_{in})
=\left(\frac{D_+(t_{in})}{D_+(t_{eq})}\right)^2\Delta(k_{in},t_{0}) .
\label{eq:DeltaT0}
\end{equation}
The evolution after horizon crossing depends on whether it occurs before or after
matter-radiation equality:
\begin{enumerate}
\item{
If $t_{in}<t_{eq}$, then
$D_{+}(t_{in})=D_{+}(t_{eq})$ because perturbations are
essentially frozen in the radiation era. Consequently, $P(k_{in})
= P(k_{eq}) \, (k_{in}/k_{eq})^{-3}$.}
\item{
If $t_{in}>t_{eq}$, then
$D_{+}(t_{in})/D_{+}(t_{eq})=(t_{in}/t_{eq})^{2/3}
=(k_{eq}/k_{in})^{2}$. In the last equality we
have used  the relation between the comoving wavenumber and the
time of horizon crossing $k_{in} \propto t_{in}^{-1/3}$, valid in the
matter era. Then $P(k_{in}) = P(k_{eq}) \, (k_{in}/k_{eq})$.}
\end{enumerate}
As a consequence of the growth of density perturbations in the
matter and radiation epochs, the power spectrum has two asymptotic
regimes: $P(k)\sim k^1, k^{-3}$ at large and small scales, respectively,
with a maximum at the scale of matter-radiation equality.
Since the transition from the radiation to the matter dominated
period is not instantaneous, $P(k)$ has a smooth
maximum about matter-radiation equality, at $k_{eq}$.
The power spectrum is conveniently expressed as $P(k)=A \, k^n\, T^2(k)$, where
$A$ is a normalization constant and $n$ the spectral index at large scales.
The transfer function $T(k)$ is determined by the growth rate
within the horizon. In the specific case of the HZ prediction, $n=1$.

\section{The matter power spectrum of generic dark energy models}\label{sec:MPS_DE}
The spectra of the concordance $\Lambda$CDM models differ from the
spectra of models with no cosmological constant in two main
respects, namely: (1) the scale of matter radiation equality is
shifted to larger scales, and (2) the growth factor of matter
density perturbations slows down once the overall expansion accelerates.
As mentioned above,
in this section we shall show how to compute the present matter power
spectrum of a DE model, in principle not implemented on a
numerical package, from the power spectrum of a model that is
implemented. As an example, we shall construct the power spectrum
of models with $w_{DE}=-0.8,-0.5,-0.1$ from the concordance
$\Lambda$CDM. The evolution of density perturbations of DE models
with constant EoS is implemented in standard packages like CMBFAST
and can be computed numerically but they will be useful to estimate
the accuracy of the method. For simplicity, all models will share identical
 cosmological parameters, namely $\Omega_\Lambda=0.73, \Omega_{DM}=0.23,
\Omega_b=0.04$, $H_0=71$ km/s/Mpc and $n=1$. The models differ only
in the  EoS parameter, $w_{DE}$.

Following the HZ prescription, we assume that all perturbations
have the same amplitude at horizon crossing. Then, we just need to
compare the growth rate of density perturbations in both models
once  perturbations cross the horizon. Even if the subhorizon
evolution of density perturbations is the same in the $\Lambda$CDM
as in the DE model, the final power spectrum could be different.
In each model,  fixed comoving wavelengths $\lambda_{in}$
cross the horizon at  different times $t_{in}$ and the growth rate
from $t_{in}$ to the present time $t_0$  will be different for
each of them. Then, we need to determine: (1) the size of the
horizon as a function of time to fix when a perturbation crosses
the horizon, and (2) solve the equations of evolution of
subhorizon density perturbations during the radiation, matter and
accelerated expansion epochs. If DM and DE density perturbations
evolve independently during the radiation regime, we can expect
the evolution of DM perturbations to be independent of the model.
Specifically, in the radiation era the expansion timescale is
$t_{exp}\propto(G\rho_{rad})^{-1/2}$ while if the free-fall time
of matter within a density perturbation is
$t_{ff}\propto(G\rho_m)^{-1/2}$, much smaller than the expansion
timescale and matter perturbations will not grow significantly
during the radiation regime. With this simplifying assumption, if
the DE model and $\Lambda$CDM have the same matter-radiation
equality and perturbations cross the horizon at the same time,
then the power spectrum at small scales will have the same shape
in both models. Without restricting the applicability of our
method, this assumption guarantees that the DE model will pass the
constraints imposed by the galaxy distribution on scales
$\lambda\le 100$Mpc$/h$ not less well than the $\Lambda$CDM model.

Our method is more easily implemented when the equations of
evolution of subhorizon sized perturbations after matter-radiation
equality form a closed system and can be solved independently for
each energy density component (see \cite{bertschinger-ma} for
extensive reviews on cosmological perturbation theory). If both
models have identical evolution during the radiation era, once the
Universe becomes matter dominated the anisotropic stress due to
neutrinos will be negligible and, in the Newtonian gauge, it will
suffice just one single gravitational potential, say $\phi$, to
determine the flat metric element
\begin{equation}
ds^2=-(1+2\phi)dt^2+a^2(1-2\phi)dx^idx_i\, .
\label{eq:metric}
\end{equation}
From the (0,0) component of Einstein's equations,
the evolution of the gravitational potential is given by
\begin{equation}
\frac{k^{2}}{a^{2}}\phi+3H\left(\dot{\phi}+H\phi\right)=
-4\pi G \sum\bar{\rho}_i\delta_i \, ,
\label{PertEinstein1}
\end{equation}
where the sum extends over all matter components. If the energy
components interact only gravitationally between themselves, then
the energy-momentum tensors are individually conserved. For a
generic component $A$, by perturbing the conservation equation
$T^{\,\, \mu \nu}_{A\,\,\, ;\mu}=0$ one obtains
\begin{eqnarray}
\label{eq:dA}
\dot{\delta}_{A}&=& -(1+w_A)\left(\frac{\theta_A}{a}-3\dot{\phi}\right)-
3H\left(c_{s,A}^{2}-w_A\right)\delta_A  ,\\
\label{eq:vA}
\dot{\theta}_A&=&-H(1-3w_A)\theta_A-\frac{\dot{w}_A}{1+w_A}\theta_A+
\frac{k^{2}}{a\left(1+w_A\right)}c_{s,A}^{2}\delta_A+\frac{k^{2}}{a}\phi  ,
\end{eqnarray}
where $c_{s,A}^{2}=\frac{\delta P_A}{\delta \rho_A}$ is the sound speed,
$\theta_A=k^2v_A$, and $v_A$ is the velocity field. For a more
general treatment, including interaction between DE and DM, see
e.g. \cite{maartens}. When the evolution is subhorizon, $k\gg aH$
and time derivatives can be neglected compared to spatial
gradients, eq.~(\ref{PertEinstein1}) reduces to the Poisson
equation:
\begin{equation}\label{Poisson}
k^{2}\phi=-4\pi G a^{2}\sum\rho_{i}\delta_{i} \, .
\end{equation}
Equations~(\ref{eq:dA}) and (\ref{eq:vA})
can be specialized to the case of DM,
baryons and DE, $w_c=w_b=c_{s,c}=c_{s,b}=0$.
If the DE sound speed is $c_{s,DE}=1$, then it will not cluster
at small scales and we can take $\delta_{DE}\approx 0$. Under this assumption,
the equations of evolution of matter density perturbations, defined as
$\bar{\rho}_m\delta_m=\bar{\rho}_c\delta_c+\bar{\rho}_b\delta_b$ are:
\begin{eqnarray}
\dot\delta_m&=&\frac{\theta_m}{a}  ,\\
\dot{\theta}_m&=&-H\theta_m-\frac{k^2\phi}{a}  ,\\
k^2\phi&=&-4\pi Ga^2\bar{\rho}_m\delta_m  .
\label{eq:m}
\end{eqnarray}
From this equations, the evolution of matter density perturbations is
described by a single second order differential equation:
\begin{equation}
\ddot{\delta}_m+2H\dot{\delta}_m-4\pi G\bar{\rho}_m\delta_m=0 .
\label{eq:delta}
\end{equation}
This equation does not depend on unknown functions, so the evolution of
matter density perturbations can be solved exactly, as required by our method.
In terms of the growth function $f\equiv\frac{d \ln\delta_{m}}{d\ln a}$,
Eq.~(\ref{eq:delta}) could be transformed into an even simpler,
first order, differential equation (see e.g. \cite{ivan-pavon,LWang}).

Let $D_+(t)$ be the growing mode solution of Eq.~(\ref{eq:delta}).
To solve the equation, we need to specify the initial conditions
at some arbitrary time. In the CMBFAST code, this is done at some
initial space-like hypersurface. In HZ, all modes have the same
amplitude at horizon crossing, i.e., amplitudes are fixed at
different times for different modes. This amplitude will depend on
the model, but for the same wavelength the ratio of amplitudes of
different models will be constant, independent of scale, and this
factor could be absorbed into the normalization. But even if the
amplitudes at horizon crossing were the same, the current
amplitudes will differ because of their different growth rates.
When a perturbation of a fixed scale $k_{in}$ enters the horizon
at time $t_{in,DE}$ and $t_{in,\Lambda CDM}$ it grows by a factor
$D_{+,DE}(t_{0,DE})/D_{+,DE}(t_{in,DE})$ and $D_{+,\Lambda
CDM}(t_{0,\Lambda CDM})/D_{+,\Lambda CDM}(t_{in,\Lambda CDM})$,
respectively. Therefore, the final amplitudes will differ by a
factor
\begin{equation}
Q(k_{in})=\frac{D_{+,DE}(t_{0,DE})/D_{+,DE}(t_{in,DE})}
{D_{+,\Lambda CDM}(t_{0,\Lambda CDM})/D_{+,\Lambda
CDM}(t_{in,\Lambda CDM})} \, . \label{eq:Q}
\end{equation}
As a result,
\begin{equation}
P_{DE}(k)=Q^2(k)P_{\Lambda CDM}(k)  .\label{eq:pk}
\end{equation}
This identity holds even if the spectral index differs from unity,
$n\ne 1$.

To derive $Q(k)$ for each cosmological model we first
compute the horizon radius, $d_H(z)$, to determine when a mode
enters the horizon, and then  solve the dynamical equations to
find the subhorizon growth factor $D_{+}(z)$ after
matter-radiation equality. If subhorizon perturbations in
$\Lambda$CDM and DE models grow at the same rate in the radiation
dominated regime, then $Q(k)=$ const and
$\sigma_{8,DE}=Q(2\pi/8h^{-1}Mpc)\sigma_{8,\Lambda CDM}$. If both
power spectra are normalized at the same amplitude at small
scales, then $Q(k\gg k_{eq})=1$ and DE and $\Lambda$CDM spectra
will coincide at small scales. The method could be applied to
superhorizon scales if the equations of evolution formed a closed
system. If not, one can simply extrapolate their amplitude using
the HZ prescription. If required, it could also be generalized to
include perturbations in other components such as DE. It suffices
 to specify the amplitude of every component at horizon
crossing and follow its subsequent evolution.

\begin{figure}[h!]
\plotone{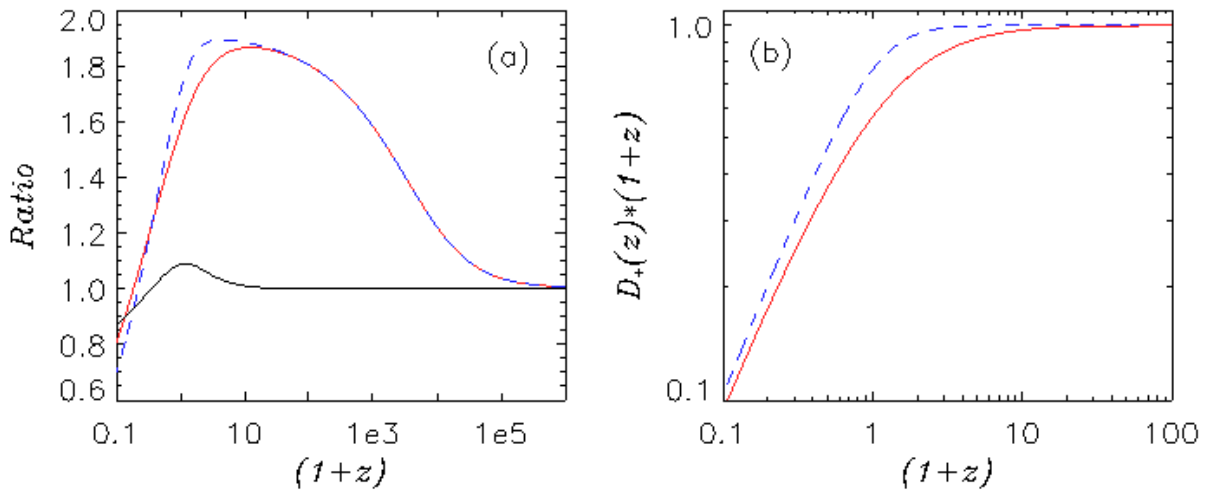} \caption{(a) Ratios of the horizon radii,
$d_{H}$,  of three cosmological models: standard CDM, concordance
$\Lambda$CDM, and the $w_{DE} = -0.5$ DE model: $d_{H,\Lambda
CDM}(z)/d_{H,CDM}(z)$ (blue dashed line),
$d_{H,DE}(z)/d_{H,CDM}(z)$ (thick red solid line), and
$d_{H,\Lambda CDM}(z)/d_{H,DE}(z)$ (thin black solid line). (b)
Growth factors of the $\Lambda$CDM (dashed blue line) and DE model
(solid red line) in units of the growth factor of the standard CDM
model, $D_{+,CDM}\sim (1+z)^{-1}$.} \label{f1}
\end{figure}

In Figure \ref{f1}a, we plot the ratio of the comoving size of the
horizon between different models; the thick solid (red) and dashed
(blue) lines correspond to the ratio of the horizon size of the
$w_{DE}=-0.5$ DE and $\Lambda$CDM models with respect to the
standard ($\Omega_\Lambda=0$, $\Omega_m=1$) CDM model. As
expected, the size of the horizon is the same in the radiation
dominated regime irrespective of the cosmology.  The thin (black)
solid line represents the ratio of the $\Lambda$CDM horizon size
to that of the $w_{DE}=-0.5$ DE model, i.e., $d_{H,\Lambda
CDM}/d_{H,DE}$. In this case the ratio is very similar up to
$z\approx 30$. The difference arises because the period of
accelerated expansion starts earlier in the DE model. In Figure
\ref{f1}b we plot the growth factor (in units of the standard CDM
growth factor) of the $w_{DE}=-0.5$ DE model, solid (red) line,
and the concordance $\Lambda$CDM model, dashed (blue) line. All
models are normalized so that the amplitudes of the growing modes
at recombination are the same: $D_+(z_{rec})=(1+z_{rec})^{-1}$,
i.e., they coincide with the growth factor of the standard CDM at
that redshift ($z_{rec} \approx 1090$).

\begin{figure}[h!]
\plotone{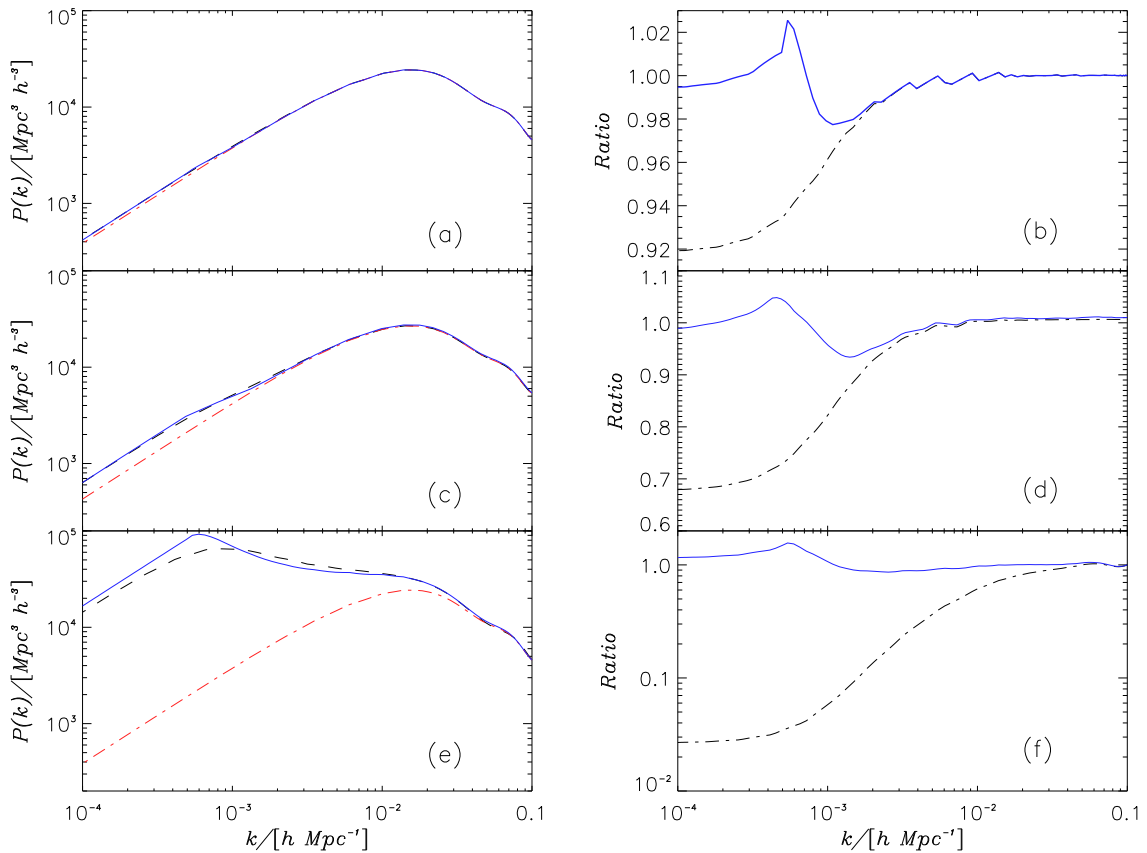} \caption{(a,c,e) Matter power spectra of the
concordance $\Lambda$CDM (dot-dashed red line) and the DE model
(solid blue line) with EoS parameter $w_{DE}=-0.8$ (a),
$w_{DE}=-0.5$ (c), and $w_{DE}=-0.1$ (e). The dashed line
correspond to the numerical (CMBFAST) solution and the (blue)
solid line corresponds to the approximated spectrum derived using
eq.~(\ref{eq:pk}). All other cosmological parameters are the same
as in the $\Lambda$CDM concordance model: $H_0 = 71$ km/s/Mpc,
$\Omega_m=0.27$, $\Omega_\Lambda=\Omega_{DE}=0.73$, and $n=1$.
(b,d,f) Ratios between different matter power spectra: $P_{\Lambda
CDM}(k)/P_{DE}(k)$ (dot-dashed black line), and
$P_{DE,approx}(k)/P_{DE}(k)$ (solid blue line). Panel (b)
corresponds to $w_{DE}=-0.8$, (d) to $w_{DE}=-0.5$, and (f) to
$w_{DE}=-0.1$.} \label{f2}
\end{figure}

In Figures \ref{f2}a,c,e we represent the exact and approximated
power spectra. In each panel, the dashed black line
represents the power spectra computed numerically using CMBFAST for
DE with $w_{DE}=-0.8$ (a), $w_{DE}=-0.5$ (c), and $w_{DE}=-0.1$
(e). Solid (blue) lines plot the power spectra derived using our
analytic approximation and dot-dashed (red) lines correspond to
the concordance $\Lambda$CDM model used to construct the
approximated solutions. For the first two models, the approximated
and exact DE spectra are almost indistinguishable on a log-log
scale. In Figures \ref{f2}b,d,f we represent the ratio of the
approximated to the exact (computed with CMBFAST) power spectrum
(solid -blue- line) for (from top to bottom)
$w_{DE}=-0.8,-0.5,-0.1$. The dot-dashed (black) line represents the ratio
of the concordance $\Lambda$CDM power spectra to the exact DE
power spectra. The accuracy of our prescription depends on the
model parameters. It is within 1-3\% for $w_{DE}=-0.8$, 1-8\% for
$w_{DE}=-0.5$, and degrades to 5-35\% for $w_{DE}=-0.1$. Properly
speaking, this latter model is not a DE model because it does not lead to
a period of accelerated expansion. Even in this extreme case and
ignoring the clustering of this ``DE", the approximation is rather
good, the largest error being 35\% at $k\sim 5\times
10^{−4}$Mpc/h. In any case, a certain discrepancy between the
numerical and approximated spectra is to be expected, as found
around $k\sim 10^{-3}$Mpc/h, since for perturbations that come
within the horizon after matter-radiation equality the
gravitational potential is still evolving with time and
Eq.~(\ref{Poisson}) becomes less accurate.

Figure.~\ref{f2} is our main result. It shows how useful our
prescription is to construct the matter power spectrum of an
arbitrary DE model. The closer the model parameters are to the matter
power spectrum used as a  starting point (in the examples above,
the $\Lambda$CDM model), the more accurate the approximation is.
Once the DE model parameters differ significantly, our approach is not
so accurate but, at the same time, the power spectrum of the DE model
separates from the concordance model. Therefore, as long as the
concordance model is a good fit to the data, the difference between
this model and the exact/approximated DE spectrum are so large
(almost a factor 30 in the $w_{DE}=-0.1$ case) that the uncertainty
of our approximation is irrelevant. Large scale structure data like CMB
temperature anisotropies on large scales would certainly rule out the
$w_{DE}=-0.1$ model, even allowing for a 50\% uncertainty in the
matter power spectrum at all scales.  The $w_{DE}=-0.5$ spectrum is
identical to $\Lambda$CDM at small scales, but  different enough
at large scales as to expect that CMB temperature anisotropies on
large angular scales could rule out the model.
For $w_{DE}=-0.8$, the approximate spectrum is so close to that
of $\Lambda$CDM that to discriminate it from the concordance model
will require background tests such as SN Ia, BAO, etc.(even if we
consider the exact matter power spectrum).
As determined by Larson {\it et al.} \cite{larson}, WMAP 7yrs
data alone yields $w_{DE}=-1.12^{+0.42}_{-0.43}$ at $1\sigma$, while
including data on BAO and high redshift supernova produces
$w_{DE}=-0.980\pm 0.053$ \cite{komatsu}. Here lies the main
advantage of our approach: one can quickly construct an
approximate power spectrum for any DE model that is more appealing
from the theoretical point of view than the $\Lambda$CDM
concordance model. When the model agrees with the data at the
background level, if the matter power spectrum is very different
from $\Lambda$CDM, it will not fit the observations of galaxy
clustering and LSS.

\section{The Radiation Power Spectrum}\label{sec:RPS}
After computing the DE power spectrum, constraints on the
model can be imposed using data on galaxy clustering and LSS. Since
the spectral shape at small scales is the same than in the $\Lambda$CDM,
once the DE model is normalized to the measured $\sigma_8$, it
will reproduce the data on small scales (galaxy clustering,
peculiar velocity amplitude, weak lensing convergence spectrum,
etc.) as well as the $\Lambda$CDM. Of the three models discussed
in Fig.~\ref{f2}, we shall restrict our analysis
to $w_{DE}=-0.5$ DE model. As shown in Figure \ref{f1},
horizon size and perturbation growth between $w_{DE}=-0.5$ DE and
$\Lambda$CDM start to differ at $z\simeq 20$ and the power
spectrum at $k\le 4\times 10^{-3}h/$Mpc (see Fig. \ref{f2}c),
scale that comes into the horizon well in the matter era. The
differences in the matter power spectrum can only be tested using
data on large scales, like CMB temperature anisotropies.

Prior to decoupling, baryon and photons are tightly coupled and
inhomogeneities in the baryon distribution are also reflected in
anisotropies on the radiation field. Several physical mechanisms
contribute to the generation of temperature anisotropies
\cite{hu-nature}.  Analytic methods that trace the structure of
the cosmic microwave background anisotropies have been used to
compute the contribution of different effects such as
gravitational redshifts, acoustic oscillations, diffusion damping,
Doppler shifts, reionization as well as the effect of curvature, a
cosmological constant and their dependence on initial conditions
\cite{hu-sugiyama}. The gravitational redshifts are the dominant
contribution on large scales. These anisotropies are strongly
dependent on the underlying matter power spectrum. At $l\le 10$,
the largest contributions come from the SW and ISW effects
\cite{sachs-wolfe}. Both components can be accurately computed in
terms of quadratures involving only the matter power spectrum
\begin{equation}
C_l^{SW}=\frac{\Omega_m^2 H_0^4}{2\pi D_+^2(0)}\int_0^\infty
k^2dk\frac{P(k)}{k^4}j_l^2(k r(z)) \, , \qquad
C_l^{ISW}=\frac{2}{\pi}\int_0^\infty k^2dkP(k)I_l^2(k)\, ,
\label{eq:cl}
\end{equation}
where $I_l(k)=3\Omega_{m,0}\frac{H_0^2}{c^2k^2}\int_0^{z_{rec}}dz
j_l(kr(z))(d[(1+z)D_+]/dz)$. In these expressions $j_l$ is the
spherical Bessel function, $H_0$ the Hubble constant,
$r(z)=\int_0^zH^{-1}(z)dz$ the look-back distance and
$\Omega_{m,0}$ the current matter density in units of the critical
density and $D_+$ is the growth factor that verifies
$D_+(z)=(1+z)^{-1}$ well in the matter dominated period,
so during that epoch there is no significant ISW effect.
The total radiation spectrum would contain contributions
such as acoustic oscillations and Doppler shifts, so $C_l\ge
C_l^{SW}+C_l^{ISW}$. The accuracy of
Eq.~(\ref{eq:cl}) could be improved by including more
contributions as discussed in \cite{atrio-doro} but,
as we will see below, the difference in the amplitude of the low
multipoles suffices to rule out the $w_{DE}=-0.5$ DE model
so the current approximation is accurate enough for our purposes.

\begin{figure}[h!]
\plotone{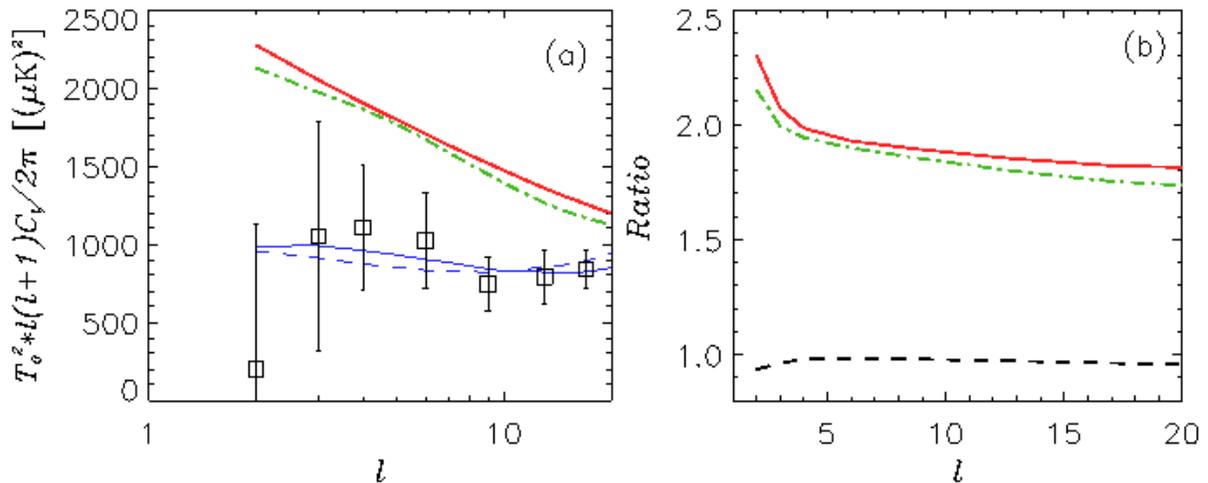} \caption{ (a) Power spectra of the CMB
temperature anisotropies. Thin dashed (blue) and solid (blue)
lines represent the exact spectrum of the $\Lambda$CDM model (with
$\sigma_{8} = 0.801$) and DE model ($\sigma_{8} = 0.53$),
respectively. Thick (red) solid and dot-dashed (green) lines plot
multipoles computed using eq.~(\ref{eq:cl}) with the exact and
approximated matter power spectrum, both normalized to $
\sigma_{8}= 0.801$. The data are WMAP 7yrs measurements. (b)
$C_{l,DE,eq~\ref{eq:cl}}/C_{l,DE,CMBFAST}$ using eq.~(\ref{eq:cl})
and the exact and approximated matter spectra, solid (red) and
dot-dashed (green) lines respectively. The dashed (black) line
represents the ratio of the previous two.}\label{f3}
\end{figure}

In Fig. \ref{f3}a, the thin (blue) solid and dashed lines
represent the radiation spectra of the DE and $\Lambda$CDM models,
respectively, computed using CMBFAST. The models are normalized
to the $\sigma_{8}$ obtained from the code,
$\sigma_{8}=0.80$ for $\Lambda$CDM and $\sigma_{8}=0.53$
for the $w_{DE}=-0.5$ DE model. The thick (red) solid line
represents the power spectra computed using Eq.~(\ref{eq:cl}) and
the dot-dashed (green) line the radiation spectra computed with
the same equation but with the approximated matter power spectra.
All matter power spectra are normalized to $\sigma_8=0.801$ \cite{larson}.
The open squares correspond to the binned power spectrum measured
by WMAP 7 yrs data \cite{lambda}; error bars include instrumental
noise and cosmic variance. To facilitate the comparison between
the different approximations, the top two lines in Fig. \ref{f3}b
represent the ratio of the DE approximated (computed using
Eq.~(\ref{eq:cl}) and the exact matter power spectrum) to the
exact DE radiation spectrum (solid red line), and the DE (using
also the approximated matter power spectrum), to the same exact DE
radiation spectrum (dot-dashed green line). The dashed (black)
line is the ratio of the previous two. The error introduced by
computing $C_l$ using Eq.~(\ref{eq:cl}) with the exact or the
approximated matter power spectrum is smaller than 10\%.

Fig.~\ref{f3} summarizes the comparison of the $w_{DE}=-0.5$ DE
model with CMB data.  When CMB anisotropies are computed using
CMBFAST and normalized to WMAP 7 yrs  data, the CMB power spectrum
fits the low multipoles rather well, but then the amplitude of
matter density fluctuations is $\sigma_{8,DE}=0.53$, well outside
the value $\sigma_8=0.801\pm 0.030$ \cite{larson} allowed by the
data. As a result, the $w_{DE}=-0.5$ DE model with parameters
$\Omega_m=0.27$, $\Omega_\Lambda=0.73$, etc. is ruled out. But if
the calculation of the CMB power spectrum proves to be difficult,
the approximated power spectrum could be used to approach the
problem differently. Once  the power spectrum of the DE model
(with $w_{DE}=-0.5$) is constructed and normalized  to
$\sigma_8=0.801$, the amplitude of temperature anisotropies at
$l\le 10$ would result $l(l+1)C_l^{DE}/2\pi\sim 2000 (\mu K)^2$,
about a factor of $2$ larger than the measured spectrum, a factor
larger than the uncertainties introduced by our approximation
(less than 10\%) or by Eq.~(\ref{eq:cl}) (less than 15\% for $l\le
10$). Therefore, without the need of further information, a model
like DE with $w_{DE}=-0.5$ could be ruled out based on the
amplitude of the low order multipoles. Models that differ from
$\Lambda$CDM also during the radiation regime can be more severely
constrained by using weak lensing, peculiar velocities, or galaxy
clustering data.

\section{Conclusions}\label{sec:Conclusions}
We have shown how to compute the matter power spectrum of DE
cosmological models using a fiducial $\Lambda$CDM model and the
growth factor of subhorizon density perturbations of the model
under consideration. This allows to use data on CMB temperature
anisotropies and galaxy clustering to discriminate models without
having to solve the evolution of density perturbations of all
matter components, thus economizing much effort. Figure \ref{f2}
shows the advantage of our proposal. In most cosmological models,
the equations describing the time evolution of matter density
perturbations can be derived from the conservation of the energy
momentum tensor and solved for each component individually. Thus,
an approximated matter power spectrum can be easily
computed. If a model fits the background evolution as determined,
for example, by luminosity distances obtained from supernovae
data, before proceeding to a detailed study of the evolution of
its density perturbations and CMB anisotropies, one can compute an
approximated matter power spectrum and predict observables such as
the low multipoles of the CMB, the ISW-LSS correlation, the weak
lensing convergence spectrum, etc., that can be compared with
observations. Even if a model agrees with the data on the
expansion rate of the Universe, it could be ruled out by data on temperature
anisotropies or galaxy clustering without requiring to solve the
first order perturbation equations in full.

Our method has its limitations: it produces an approximated matter
power spectrum and only observables derived from it can be used.
Since it fixes amplitudes at horizon crossing, it is insensitive
to any instability that could occur on superhorizon scales such as
those present in some models with DM/DE interactions
\cite{tegmark}. However, as our examples show, the
approximated and numerical power spectra are very similar when the
model parameters are close to those of the starting model.  This
is the advantage of our approach: the approximated power spectrum
can be used to distinguishing models, that despite reproducing the
background observational data, would fail to fit the galaxy
clustering and LSS data.  In this respect, the $w_{DE}=-0.8$ model
is so close to $\Lambda$CDM it should be tested with background data
such as SN Ia, BAO etc. to be ruled out (the same happens with the
exact matter power spectrum). In the  $w_{DE}=-0.5$ DE model, the
relative error in the approximated and numerical $P(k)$ is smaller
than 8\%, but the difference with the concordance model is so
significant that the predicted CMB temperature anisotropies on
large scales are about a factor of 2 larger than the measurements,
a factor much larger than the uncertainty introduced by our
approximation or by Eq.~(\ref{eq:cl}). Finally, the $w_{DE}=-0.1$
is so different from $\Lambda$CDM that it can be ruled out at a glance,
irrespectively of the poorer quality of our approximation. When the observables
derived from the approximated matter power spectrum fails to fit
the data, we can confidently expect the model to fail. But even if
a model fits some data on galaxy clustering, it may not
necessarily reproduce all the observations, like the full spectrum
of CMB temperature anisotropies; it simply means that it deserves
further study.

\acknowledgments {The research of I.D. and D.P. was partially
supported by the ``Ministerio Espa\~{n}ol de Educaci\'{o}n y
Ciencia" under Grant No. FIS2009-13370-C02-01 and by the
``Direcci\'{o} de Recerca de la Generalitat'' under Grant Number
2009SGR-00164. I.D. was funded by the ``Univesitat Aut\`{o}noma de
Barcelona" through a PIF fellowship. F.A-B. acknowledges financial
support from the spanish Ministerio de ``Educaci\'on y Ciencia",
Grants No. FIS2009-07238 and CSD 2007-00050}.

\end{document}